\documentclass[runningheads]{llncs}
\pdfoutput=1
\usepackage{graphicx}
\usepackage{float}
\usepackage{hyperref}
\usepackage{booktabs}
\usepackage{tablefootnote}
\usepackage[utf8]{inputenc}
\usepackage[backend=biber,sorting=nyt,bibencoding=ascii,abbreviate=false,mincrossrefs=3,style=numeric,maxnames=1]{biblatex}
\newcommand{\textcode}[1]{\texttt{\small{#1}}}

\begin{document}
\title{The Water Health Open Knowledge Graph}
%
%
\author{Gianluca Carletti\inst{3} \and Elio Giulianelli\inst{4}\orcidID{0000-0003-0998-9751} \and \\ Anna Sofia Lippolis\inst{1,2}\orcidID{0000-0002-0266-3452}\thanks{Corresponding author. PhD student at the University of Bologna.} \and
Giorgia Lodi\inst{1}\orcidID{0000-0001-6020-5874} \and
Andrea Giovanni Nuzzolese\inst{1}\orcidID{0000-0003-2928-9496} \and \\ Marco Picone\inst{4}\orcidID{0000-0003-0126-7914} \and Giulio Settanta\inst{4}\orcidID{0000-0002-6454-2859}\thanks{The authors are sorted alphabetically as they equally contributed to this paper.}}
\authorrunning{A. S. Lippolis et al.}
%
\institute{
CNR Institute of Cognitive Sciences and Technologies, Rome, Italy \and
University of Bologna, Italy
\\
\email{\{annasofia.lippolis,giorgia.lodi,andrea.nuzzolese\}@istc.cnr.it} 
\and ARIA SpA, Milan, Italy \\
\email{gianluca.carletti@ariaspa.it}
\and Italian National Institute for Environmental Protection and Research, Rome, Italy \\
\email{\{elio.giulianelli,marco.picone,giulio.settanta\}@isprambiente.it} 
}
\maketitle              
\begin{abstract}

Recently, an increasing interest in the management of water and health resources has been recorded. This interest is fed by the global sustainability challenges posed to the humanity that have water scarcity and quality at their core. 
Thus, the availability of effective, meaningful and open data is crucial to address those issues in the broader context of the Sustainable Development Goals of clean water and sanitation as targeted by the United Nations.
In this paper, we present the Water Health Open Knowledge Graph (WHOW-KG) along with its design methodology and analysis on impact. WHOW-KG is a semantic knowledge graph that models data on water consumption, pollution, infectious disease rates and drug distribution. The WHOW-KG is developed in the context of the EU-funded WHOW (Water Health Open Knowledge) project and aims at supporting a wide range of applications: from knowledge discovery to decision-making, making it a valuable resource for researchers, policymakers, and practitioners in the water and health domains. The WHOW-KG consists of a network of five ontologies and related linked open data, modelled according to those ontologies. 


\end{abstract}
\section{Introduction}
\label{sec:intro}
Interest in water and sanitation management has grown in recent years driven by global sustainability challenges that prioritise, among the others, clean water and sanitation, as outlined in the UN Sustainable Development Goals\footnote{https://sdgs.un.org/goals.}.

To provide effective responses to these global issues, the availability of high quality and open data becomes an essential requirement. However, the heterogeneity and complexity of water and health data, when available, can pose significant challenges. Not only data is heterogeneous both in format and in semantics, but mostly it does not guarantee FAIR at any level: it is not findable, thus it is not accessible nor interoperable. There may also be no licenses specified for enabling a direct reuse of the data. In response, only a few ontological modelling solutions have emerged to represent this fragmented knowledge within a FAIR framework, aiming to cater to the need for coverage of heterogeneous datasets in the international landscape.

This paper introduces the Water Health Open Knowledge Graph (WHOW-KG), which is the first European open distributed knowledge graph aimed at linking, using a common semantics, data on water consumption and quality with health parameters (e.g., infectious diseases rates, general health conditions of the population). Designed to understand the impact of water-related climate events, water quality, and water consumption on health, it provides a harmonized data layer that can be re-used for analysis, research, and development of innovative services and applications. The project's primary driver was to establish a sustainable methodology for open knowledge graph production to ensure authoritativeness, timeliness, semantic accuracy, and consistency data quality characteristics, as well as metadata compliance with the European DCAT-AP profile\footnote{\url{https://joinup.ec.europa.eu/collection/semantic-interoperability-community-semic/solution/dcat-application-profile-data-portals-europe/release/11}.} and related national and thematic extensions.

The WHOW-KG is still under development and currently consists of more than 100 millions triples from 19 selected datasets according to three use cases. The WHOW-KG is distributed and it is available via three SPARQL endpoints: two endpoints available from two data providers (Lombardy Region and Italian National Institute of Environmental Research (ISPRA)) and one endpoint from the Institute of Cognitive Sciences and Technologies of CNR (ISTC-CNR). All the resources from Lombardy Region are licensed under the Creative Commons Public Domain License (CC0) and the ones from ISPRA under the Creative Commons Attribution 4.0 International (CC-BY 4.0) license.

In summary, this paper presents the following contributions:
\begin{itemize}
  \item The WHOW-KG and an analysis of both its impact and its impact results.
  \item A design methodology to support data providers' publication of Linked Open Data that is highly extensible and sustainable.
  \item An analysis of the five WHOW ontologies, including a review of the state of the art in terms similar works in both domains water and health.
\end{itemize}

The rest of the paper is organized as follows: (i) Section~\ref{sec:method} discusses the design methodology; (ii) Section 3 addresses the results achieved in terms of the ontology network and produced Linked Open Data; (iii) Section~\ref{sec:evalimpact} discusses the impact of the WHOW-KG; (iv) Section~\ref{sec:related} presents the related work; finally, (v) Section~\ref{sec:conlusions} concludes the paper, discusses the limitations, and defines future directions of research.

\section{Material and method}
\label{sec:method}
The WHOW-KG was developed to cope with three selected use cases identified in the context of the WHOW project with domain experts and data providers. Those use cases are: (i) Contaminants in marine waters (UC1), (ii) Water quality and for human consumption (UC2), and (iii) Meteorological extreme events (UC3). The first use case, i.e.  Contaminants in marine waters, aims at modelling ontologies and creating linked open data on human exposure to chemicals and biological contaminants in marine waters, ingestion of contaminated fish products, and airborne
exposure, such as Ostreopsis Ovata\footnote{Ostreopsis Ovata is a well known genus of free-living dinoflagellates found in marine environments that is frequently associated with phenomena of human intoxication.}. The second use case, i.e. Water quality and for human consumption, focuses on generating ontologies and linked open data for modelling and representing quality of surface and ground waters as well as drinking water quality parameters and values, measured by compliance with the EU Directive 2020/2184\footnote{\url{https://eur-lex.europa.eu/eli/dir/2020/2184/oj}.} on the quality of water intended for human consumption. Finally, the third use, i.e., Meteorological Extreme events, is about modelling ontologies and linked open data for representing meteorological phenomena, alteration of the hydrological cycle and agriculture industries. More details about the use cases can be found by interested readers in a public deliverable~\cite{Picone2021} of the project.

\subsection{Material}
\label{sec:material}
The aforementioned use cases were defined along with the identification of pertinent core open datasets by means of a co-creation programme organised within the scope of the WHOW project. Hence, 77 participants actively contributed to the programme. The group of co-creators included domain experts, stakeholders, practitioners, and data providers from both public and private organisations located in the EU. The full list of datasets identified can be consulted in a corresponding project deliverable~\cite{Carletti2023}. From this list, we selected high-priority datasets that are currently used for designing and generating the WHOW-KG. The selection was done in compliance with to the following criteria: (i) relevance to the use cases; (ii) open licence associated with the dataset; and (iii) data availability for the years spanning mainly from 2018 to 2021. For some datasets, the time period is even longer starting from 1999 to 2023. In general, time span of reference 2018-2021 is a requirement defined in the project and contributors of the co-creation programme. The identified datasets are reported in Table~\ref{tab:datasets} with an identifier, short description, data format, right holder, supported use case, and number of records. By number of records we mean the number of rows and triples for CSV and RDF data sources, respectively.

\begin{table}[ht!]
\begin{center}
	\caption{Datesets selected for the creation of the WHOW-KG.}
	\label{tab:datasets}
        \resizebox{.9\textwidth}{!}{
        \begin{tabular}{p{0.5cm}p{3.5cm}clcr}
	\toprule
	{ \bf ID} & { \bf Description } & {\bf Format} & {\bf Right holder} & {\bf Use Case} & {\bf \# of records}\\
	\midrule
	D1\tablefootnote{\url{https://www.dati.lombardia.it/Ambiente/Dati-analitici-corpi-idrici-fluviali/kr6i-f553}.} & Analytical data of river water bodies, including flow rate & CSV & ARPA Lombardia & UC2 & 1 060 320\\
       D2\tablefootnote{\url{https://www.dati.lombardia.it/Ambiente/Dati-analitici-corpi-idrici-lacustri/d4ep-yvbw}.} & Analytical data of lake water bodies & CSV & ARPA Lombardia & UC2 & 136 085 \\
       D3\tablefootnote{\url{https://www.dati.lombardia.it/Ambiente/Dati-analitici-acque-sotterranee/46wy-4ydd}.} & Analytical data of groundwater & CSV & ARPA Lombardia & UC2 & 591 389 \\
       D4\tablefootnote{\url{https://dati.lombardia.it/Ambiente/Altezza-laghi/xiye-qjzy}.} & Height of the lakes & CSV & ARPA Lombardia & UC2 & 5480 \\
       D5\tablefootnote{\url{https://www.dati.lombardia.it/d/fvk5-jiuq}.} & Infectious diseases rates by sex and age & CSV & Regione Lombardia & UC2 & 11 435 \\
       D6\tablefootnote{\url{https://github.com/whow-project/datasets/blob/main/RML-RULES/ostreoptis-monitoring/Ostreopsis_Ovata_AllRegions_withISTATcode_withSeas.csv}.} & Ostreopsis ovata & CSV & ISPRA & UC1 & 1,222 \\
       D7\tablefootnote{\url{http://dati.isprambiente.it/downloads/dissesto.nt.gz}.} & Repertory of mitigation measures for National Soil Protection & RDF & ISPRA & UC3 & 1 286 758 \\
       D8\tablefootnote{\url{http://dati.isprambiente.it/downloads/soiluse.nt.gz}.} & Soil consumption indicators & RDF & ISPRA & UC3 & 1 625 802 \\
       D9\tablefootnote{\url{https://github.com/whow-project/datasets/tree/main/RML-RULES/weather-mapping}.} & Meteo observations and weather stations (for october 2019 of 8 geographical Lombardy areas) & CSV & ARPA Lombardia & UC3 & 4 627 443 \\
	\bottomrule
	\end{tabular}
        }
        \end{center}
\end{table}

\subsection{Method}
\label{sec:methodology}
The methodology we used for constructing the WHOW-KG is inspired by the one defined in~\cite{Carriero2021} and relies on eXtreme Design~\cite{Blomqvist2016} (XD) for ontology modelling. XD emphasises the reuse of ontology design patterns~\cite{Gangemi2009} (ODPs) into an iterative and incremental process. More interestingly, XD is a collaborative methodology that fosters the cooperation among multiple actors with different roles (e.g. knowledge engineers, domain experts, etc.) to make sure all the modelling requirements are first captured and then effectively covered. Hence, we opted for XD since it fits our collaborative setting based on the co-creation programme. Furthermore, there is evidence il literature~\cite{Blomqvist2010} that the reuse of ODPs (i) speeds up the ontology design process, (ii) eases design choices, (iii) produces more effective results in terms of ontology quality, and (iv) boosts interoperability. 


\begin{figure}[!hbt]
\centering
\includegraphics[width=0.8\textwidth]{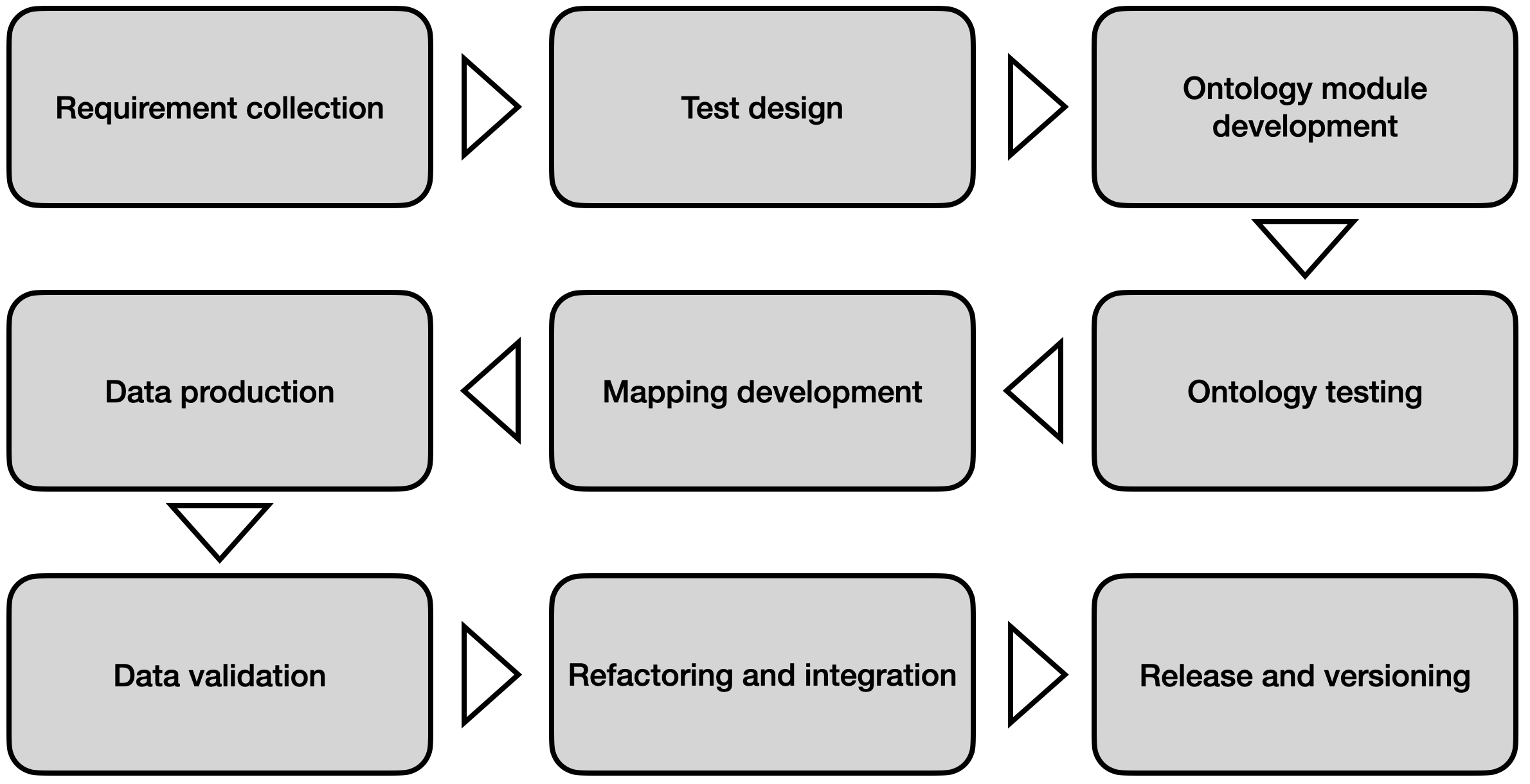}
\caption{Methodology implemented for constructing the WHOW-KG.}
\label{fig:methodology}
\end{figure}

Figure~\ref{fig:methodology} shows the methodology implemented for constructing the WHOW-KG. 
\paragraph{\textbf{Ontology design.}}
In such a figure, the activities named {\em requirement collection}, {\em test design}, {\em ontology module development}, and {\em ontology testing} come from XD and focus on ontology design. The requirement collection activity aims at eliciting the requirements as {\em competency questions}~\cite{Gruninger1995} (CQs). CQs are natural language questions conveying the ontological commitment expected from a knowledge graph (KG) and drive both ontology modelling and validation. In fact, on the one hand CQs are a means for ontology development. On the other, they can be converted to formal queries in order to assess the effectiveness of the resulting KG to cope with the requirements. We implemented the validation into the {\em ontology testing} activity. This was done by converting defined CQs into SPARQL and executing the latter as unit tests with toy data following the solution defined in~\cite{Carriero2019}. The ontology development we applied is modular (cf. activity named {ontology module development}) allowing us to generate a set of networked ontologies. Each ontology of the network is a separate module designed with the purpose of minimising coupling with other ontology modules and maximising the internal cohesion of its conceptualisation. The re-use of external ontologies and ODPs was done by applying both the direct and indirect approach~\cite{Presutti2016,Carriero2020}. Direct re-use is about embedding individual entities or importing implementations of ODPs or other ontologies in the network, thus making it highly dependent on them. Instead, indirect re-use is about applying relevant entities and patterns from external ontologies as templates, by reproducing them in the ontologies of the network and providing possible extensions. We opted for direct re-use in case of widely adopted vocabularies, such as SKOS, the Time ontology available in the Italian national catalog of semantic assets for public administrations\footnote{https://schema.gov.it}, aligned with the W3C time ontology, and the top-level\footnote{\url{https://github.com/whow-project/semantic-assets/blob/main/ispra-ontology-network/top/latest/top.rdf}.} (TOP) and environmental monitoring facilities\footnote{\url{https://github.com/whow-project/semantic-assets/blob/main/ispra-ontology-network/inspire-mf/latest/inspire-mf.rdf}.} (EMF) ontologies of the Linked ISPRA project\footnote{https://dati.isprambiente.it/}. TOP is used as a top-level ontology that provides general concepts and relations, whilst EMF provides core domain concepts and relations for modelling environmental monitoring data. On the contrary, we opted for the indirect approach for re-using patterns and to support interoperability with other pertinent ontologies, e.g. SSN/SOSA\footnote{\url{https://www.w3.org/TR/vocab-ssn/}.} \cite{Haller2019}. The latter case was realised by means of alignments axioms, such as rdfs:subClassOf and owl:equivalentClass in dedicated alignment ontologies.

\paragraph{\textbf{Linked Open Data production.}}
Once the ontology network is modelled the next steps in the methodology aims at populating the KG with Linked Open Data (LOD) gathered from the identified input data sources (cf. Table~\ref{tab:datasets}). The LOD production was performed by means of declarative mappings. Hence, in the activity {\em mapping development} we defined those mappings by means of the RDF Mapping Language\footnote{\url{https://rml.io/specs/rml/}.}~\cite{Dimou2014} (RML), which extends the W3C-standardised mapping language R2RML\footnote{\url{https://www.w3.org/TR/r2rml/}.} for mapping to RDF kind of structured data source. All the RML mapping rules defined are available on the project's GitHub repository\footnote{\url{https://github.com/whow-project/datasets}.}. These mappings were processed with both RMLMapper\footnote{https://github.com/RMLio/rmlmapper-java} and pyRML\footnote{\url{https://github.com/anuzzolese/pyrml/}.}. The latter is a lightweight Python engine for processing RML files designed and implemented in the context of the project. Data validation was then performed by using the same SPARQL unit tests derived from CQs. We point that in the latter case the unit tests were executed on real data. The activities related to data production are meant to be executed in a decentralised and distributed fashion in which different data providers might use their data and RML mapping rules independently.

\section{Results}
\label{sec:results}
\subsection{Ontology Network}
\label{sec:ontonet}

\begin{figure}[!hbt]
\centering
\includegraphics[width=.8\textwidth]{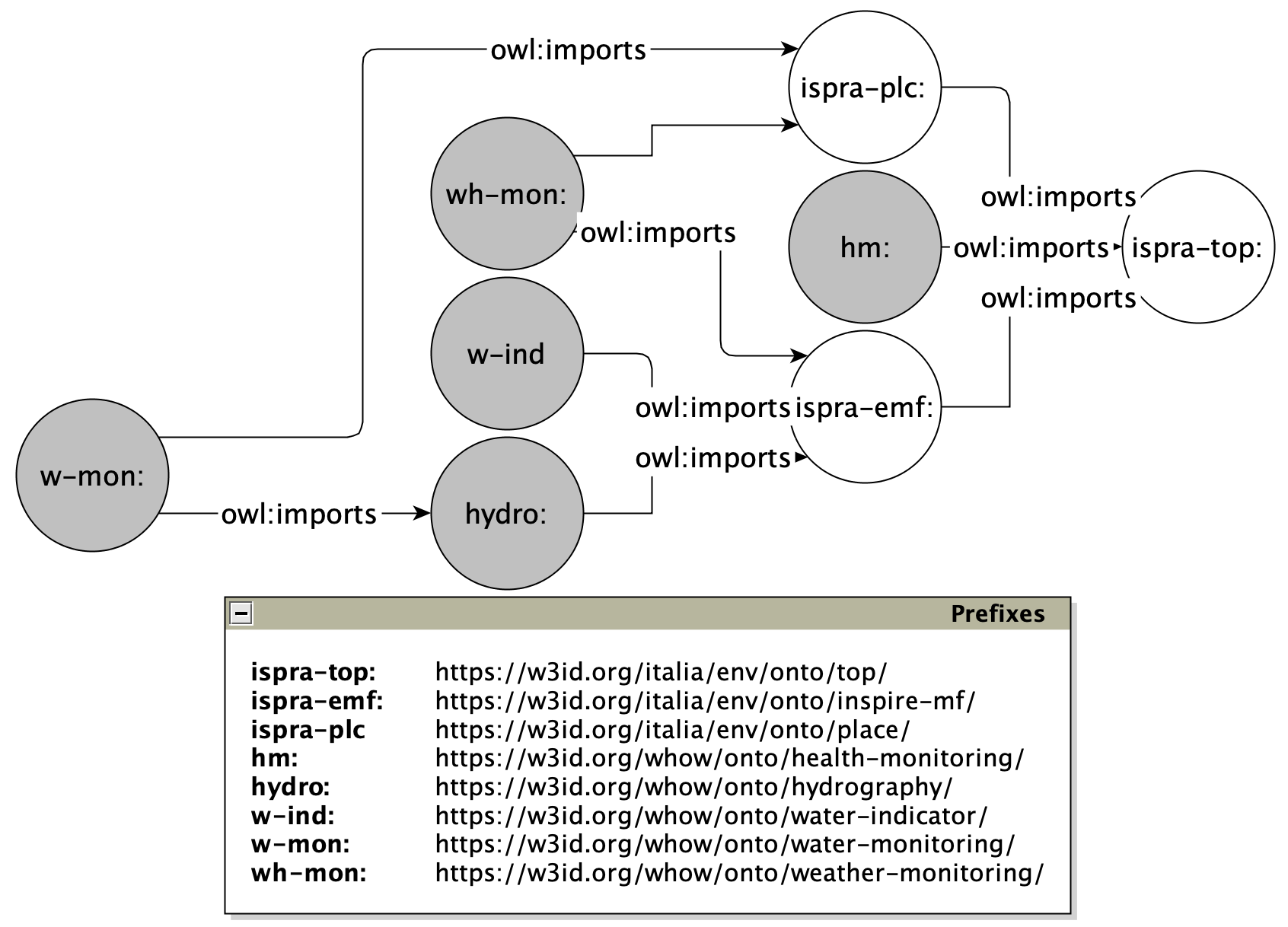} 
\caption{The WHOW ontology network.}
\label{fig:ontonet}
\end{figure}

The WHOW ontology network consists of 8 ontology modules. In Figure~\ref{fig:ontonet} each ontology is represented as a circle, whilst the arrows represent \textcode{owl:imports} axioms among the ontologies. The ontologies represented as white circles are external ontologies we re-used with the direct approach. The ontologies represented as gray circles are the novel contributions. The base namespace defined novel ontologies is \textcode{https://w3id.org/italia/whow/onto/}. From this base namespace each module defines its local namespace following the table of prefixes reported in Figure~\ref{fig:ontonet}. Table~\ref{tab:ontonet-stats} reports core metrics about the ontology network, which is: (i) under version control on GitHub\footnote{\url{https://github.com/whow-project/semantic-assets/tree/main/ontologies}.}; (ii) shared on Zenodo\footnote{\url{https://doi.org/10.5281/zenodo.7916179}.} with a CC-BY 4.0 International licence; and (iii) findable on Linked Open Vocabularies\footnote{\url{https://lov.linkeddata.es/dataset/lov/}.}.

\begin{table}[ht!]
\begin{center}
	\caption{Statics of the ontology network.}
	\label{tab:ontonet-stats}
	\resizebox{.7\textwidth}{!}{
        \begin{tabular}{lrclr}
	\toprule
	{ \bf Metric} & { \bf Value } & & { \bf Metric} & { \bf Value } \\
	\midrule
	Axioms & 2,672 & &  SubObjectPropertyOf axioms & 137 \\
        Classes & 120 & &  Inverse object properties & 61 \\
        Object properties & 161 &  & Transitive object properties & 10 \\
        Datatype properties & 21 &  & Declared property domains & 155 \\
        DL expressivity & SRIQ(D) & &  Declared property ranges & 153 \\
        SubClassOf axioms & 255	& &  Property chains & 6 \\
        Disjoint classes & 22 & &  Annotation assertions & 1,412 \\
        \bottomrule
	\end{tabular}
        }
        \end{center}
\end{table}

\begin{figure}
\centering
\includegraphics[width=\textwidth]{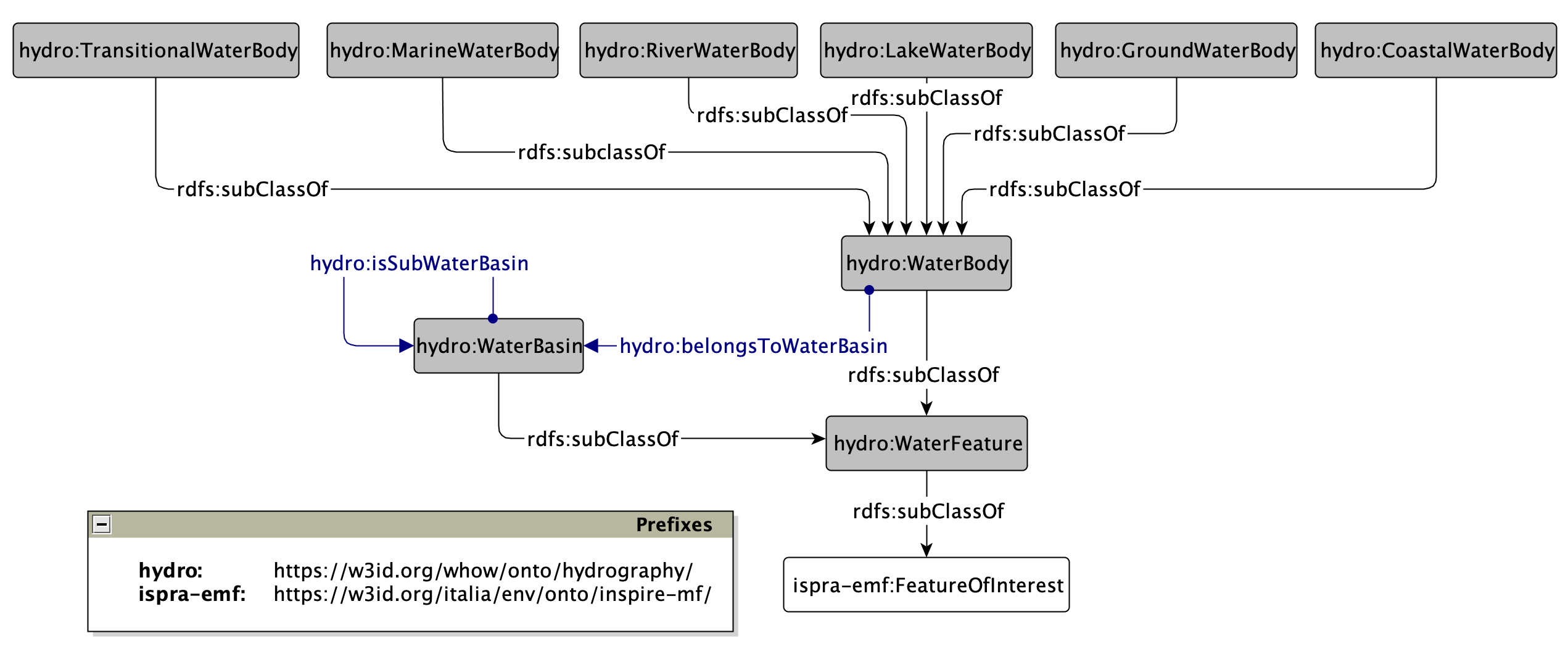} 
\caption{The Hydrography ontology.}
\label{fig:hydro}
\end{figure}

\paragraph{\textbf{Hydrography module.}}
The {\em Hydrography} ontology (prefix \textcode{hydro:}\footnote{The prefix \textcode{hydro:} stands for the namespace \url{https://w3id.org/whow/onto/hydrography}.}) represents a general-purpose hydrological taxonomy following the definitions given in the European Directive 2000/60/EC\footnote{\url{https://eur-lex.europa.eu/legal-content/EN/TXT/HTML/?uri=CELEX:32000L0060\&rid=2}.}. The \textcode{hydro:} ontology is depicted in Figure~\ref{fig:hydro} using Graffoo as reference notation~\cite{Falco2014}. With white rectangles we indicate classes directly re-used from external ontologies and with grey rectangles new defined classes.
The top-level class is \textcode{hydro:WaterFeature}, a subclass of the ISPRA ontology \textcode{ispra-emf:FeatureOfInterest} with \textcode{hydro:WaterBasin} and \textcode{hydro:WaterBody} as subclasses. A \textcode{hydro:WaterBody} further specialises into a number of subclasses defining a clear classification among the different types of water bodies. Those subclasses are \textcode{hydro:TransitionalWaterBody}, \textcode{hydro:MarineWaterBody}, \textcode{hydro:RiverWaterBody}, \textcode{hydro:LakeWaterBody}, \textcode{hydro:GroundWaterBody}, and \textcode{hydro:CoastalWaterBody}. In this ontology we reused the PartOf ODP\footnote{\url{http://ontologydesignpatterns.org/wiki/Submissions:PartOf}.} for expressing parthood between water basins (cf. the object property \textcode{hydro:isSubWaterBasin}).

\paragraph{\textbf{Water Monitoring module.}}

\begin{figure}[!hbt]
\includegraphics[width=1.2\textwidth]{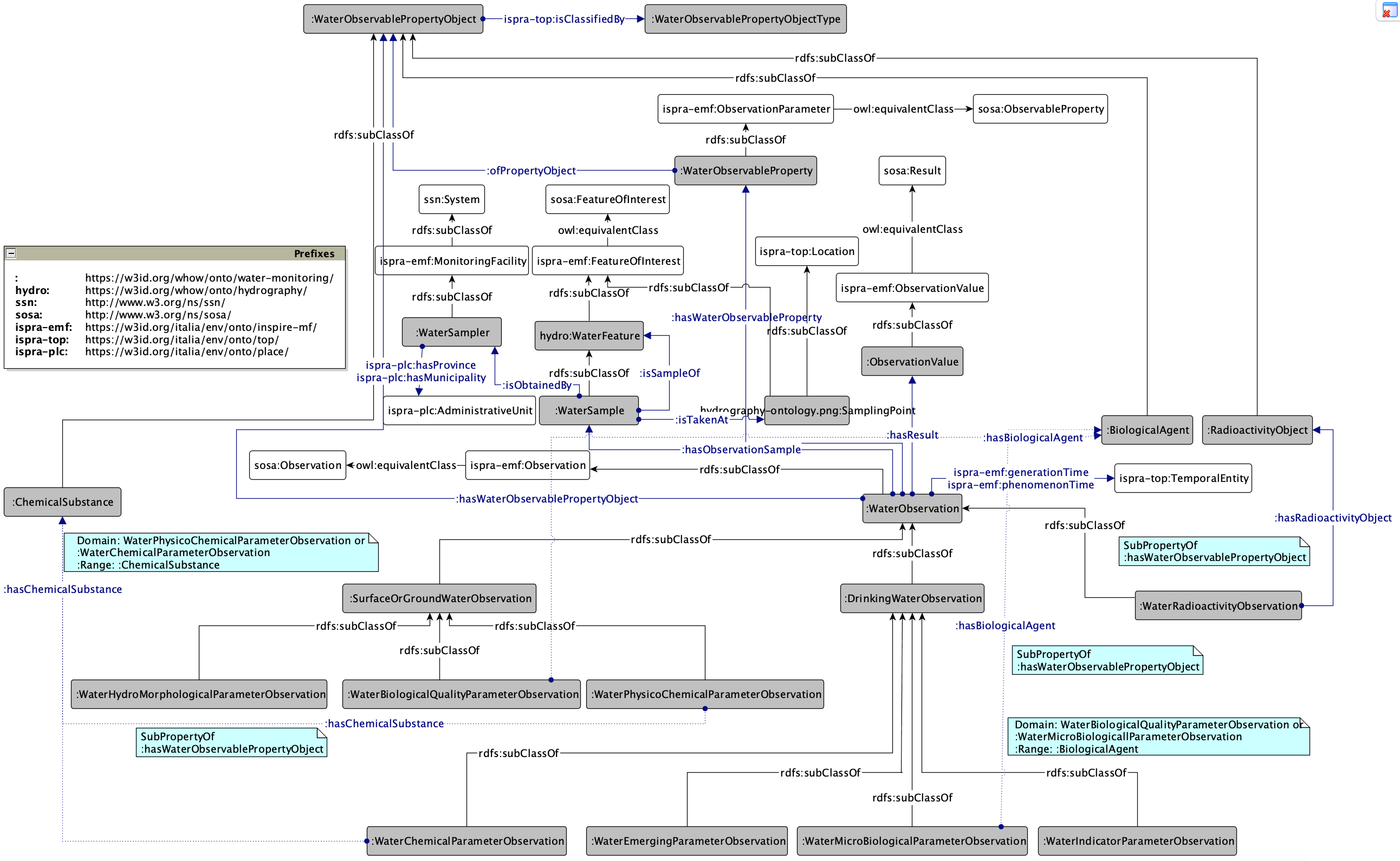} 
\caption{The Water Monitoring ontology.}
\label{fig:w-mon}
\end{figure}

The {\em Water Monitoring} ontology is identified by the prefix \texttt{w-mon:}\footnote{The prefix \textcode{w-mon:} stands for the namespace \url{https://w3id.org/whow/onto/water-monitoring}.}. It provides means to represent observations related to the quality of water courses, such as chemical and biological substances found in water bodies. The requirements for the representation of water observations are defined according to the data provided by the data providers involved in the project and the standards and directives in terms of observations and water-related assessments.  
For what concerns the representation of water observations, it is possible to refer to European directives: (i) those deriving from taxonomies from European Directive 98/83/CE (and subsequent ones)\footnote{\url{https://eur-lex.europa.eu/legal-content/EN/ALL/?uri=CELEX:31998L0083}.}, confirmed by the Italian Ministry of Health\footnote{Water quality parameters published by Italian Ministry of Health: \url{https://www.salute.gov.it/portale/temi/p2_6.jsp?lingua=italiano&id=4464&area=acque_potabili&menu=co}.}, concerning parameters of the waters for human consumption, and (ii) those deriving from the European Directive 2009/90/EC\footnote{\url{https://eur-lex.europa.eu/legal-content/EN/TXT/HTML/?uri=CELEX:32000L0060&rid=2}.}, concerning parameters of surface waters. Thus, water quality monitoring requires the integration of heterogeneous types of both observations and observation objects derived from samplers. As a result, in the ontology (cf. Figure~\ref{fig:w-mon}), a \textcode{w-mon:WaterObservation} is divided into \textcode{w-mon:DrinkingWaterObservation}, \textcode{w-mon: SurfaceOrGroundWaterObservation}, and \textcode{w-mon:RadioActivityObservation}, which are, in turn, further divided into subclasses based on the specific parameter being observed. In fact, the observations that have as an object a microbiological agent or a chemical substance, monitor it through its concentration in the water. On the contrary, observations on properties of water, such as hardness, density or pH, do not imply the presence of an object being observed sinse no chemical substance or microbiological agent is implied there. The ontology follows the Stimulus-Sensor-Observation Ontology Design Pattern (SSO ODP)~\cite{SSOPattern}, which is a standard for the Infrastructure for Spatial Information in Europe~\cite{inspire}, and the Specimen model of ISO 19156:2011\footnote{\url{http://www.iso.org/iso/catalogue_detail.htm?csnumber=32574}.}, which outlines the properties of sampling process features.

\begin{figure}[!hbt]
\centering
\includegraphics[width=\textwidth]{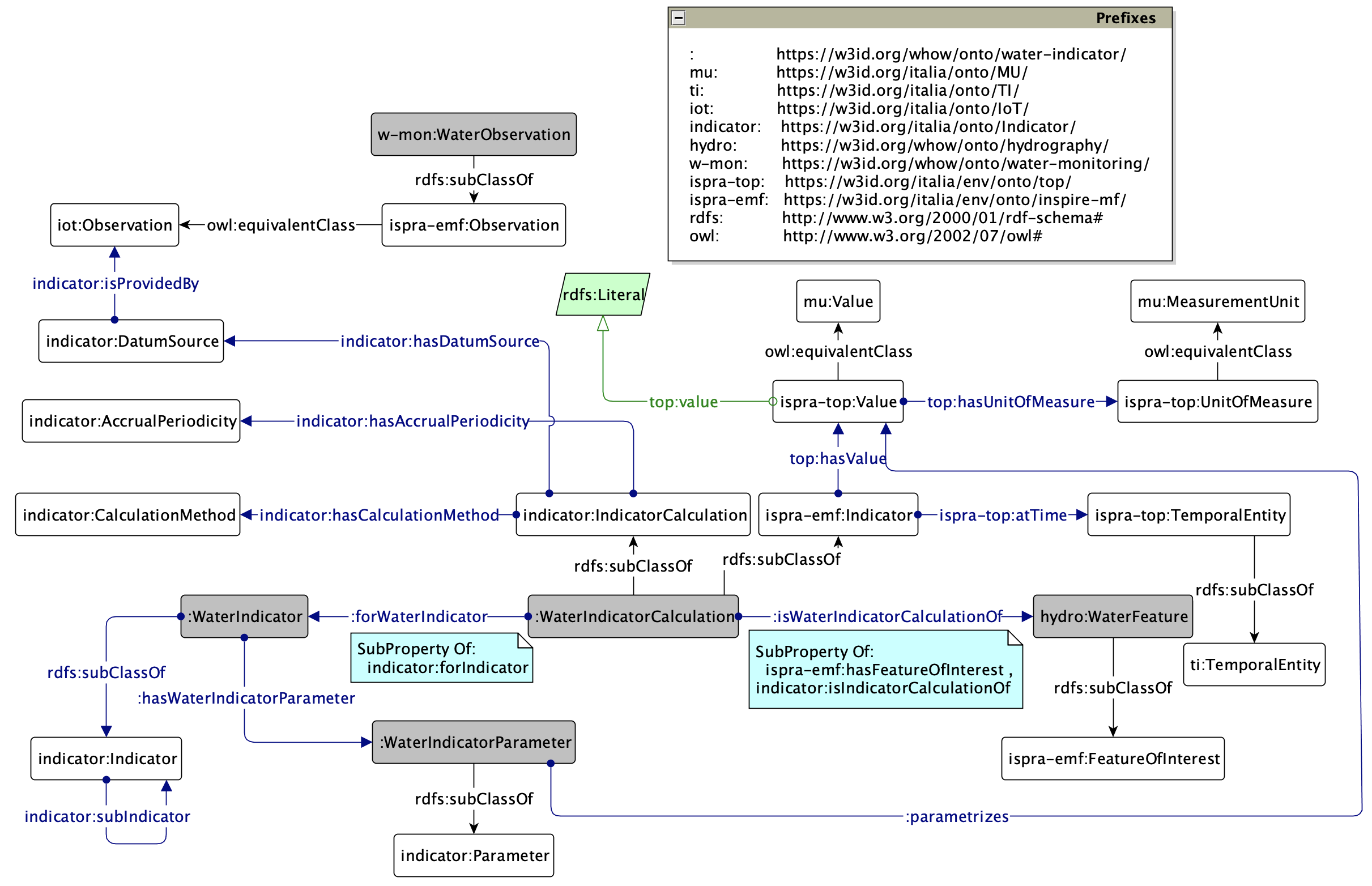} 
\caption{The Water Indicator ontology.}
\label{fig:w-ind}
\end{figure}

\paragraph{\textbf{Water Indicator module.}}
The \textbf{Water Indicator} ontology, with prefix \textcode{w-ind:}\footnote{The prefix \textcode{w-ind:} stands for the namespace \url{https://w3id.org/whow/onto/water-indicator}.}, re-uses the Indicator ontology design pattern\footnote{\url{https://github.com/italia/daf-ontologie-vocabolari-controllati/tree/master/Ontologie/Indicator/latest}.} defined in OntoPiA\footnote{\url{https://github.com/italia/daf-ontologie-vocabolari-controllati/tree/master}.}, which is the Italian national network of ontologies and controlled vocabularies. This pattern is re-used to address indicators and metrics for the indicator calculation of water quality. As shown in Figure~\ref{fig:w-ind}, the indicators can be bathing water quality classes or indicators of lakes' chemical status.

\begin{figure}[!hbt]
\centering
\includegraphics[width=\textwidth]{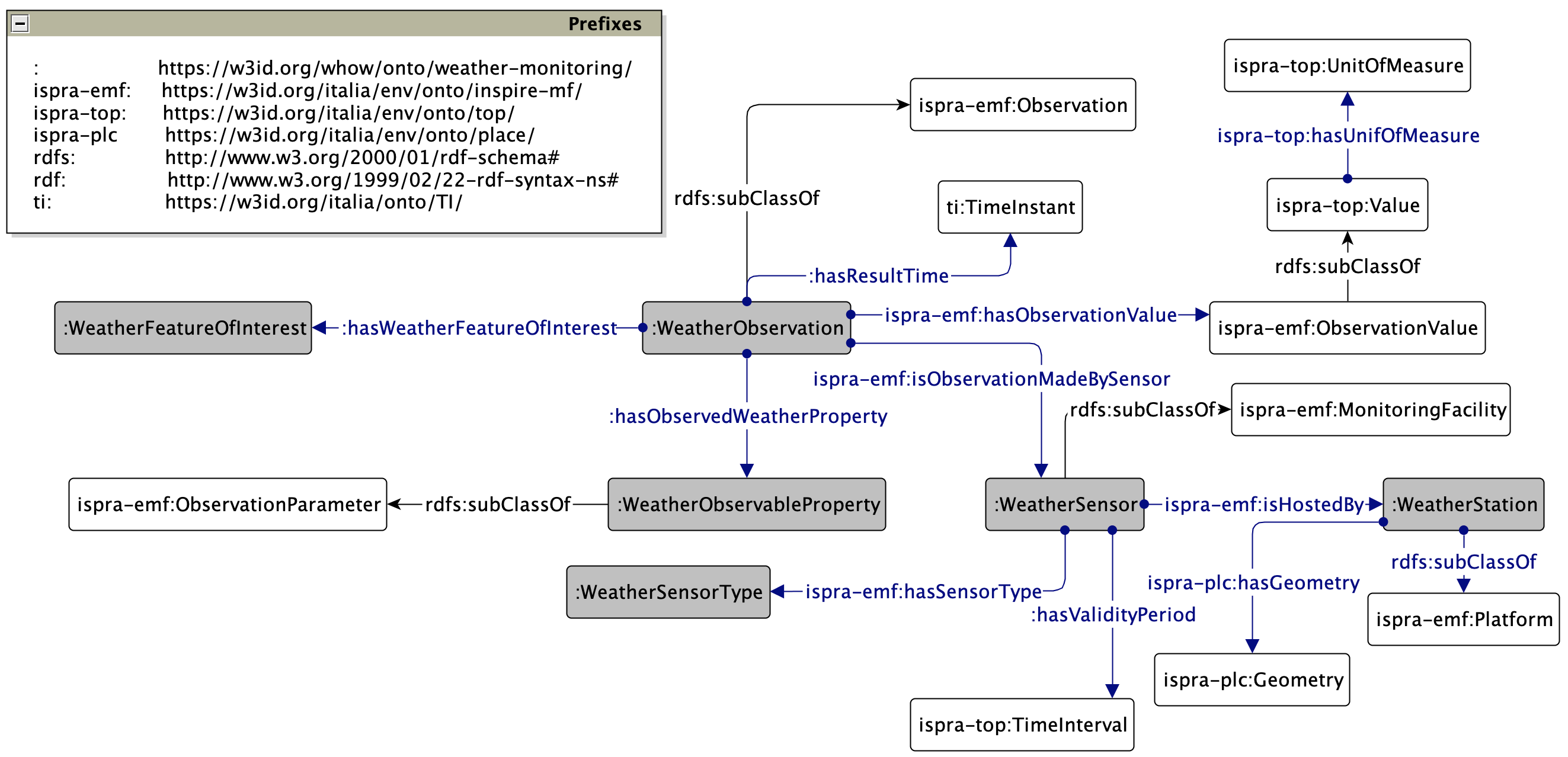} 
\caption{The Weather Monitoring ontology.}
\label{fig:wh-mon}
\end{figure}

\paragraph{\textbf{Weather Monitoring module.}}
Similarly to the Water Monitoring module, the {\em Weather Monitoring} ontology, with prefix \textcode{wh-mon:}\footnote{The prefix \textcode{wh-mon:} stands for the namespace \url{https://w3id.org/whow/onto/weather-monitoring}.} (cf. Figure \ref{fig:wh-mon}), has its focus on a \textcode{wh-mon:WeatherObservation} related to a \textcode{wh-mon:WeatherFeatureOfInterest} (either ground-level soil, air, wind, snow or rainfall), \textcode{wh-mon:WeatherObservableProperty} and \textcode{wh-mon:WeatherSensor} hosted by a \textcode{wh-mon:WeatherStation}. It reuses the ISPRA ontology network to model observations and related properties. This model is meant to address the need to represent weather observations that could serve as a basis to derive information on extreme events monitoring and prediction, such as rainfalls and snow levels.

\begin{figure}
\centering
\includegraphics[width=\textwidth]{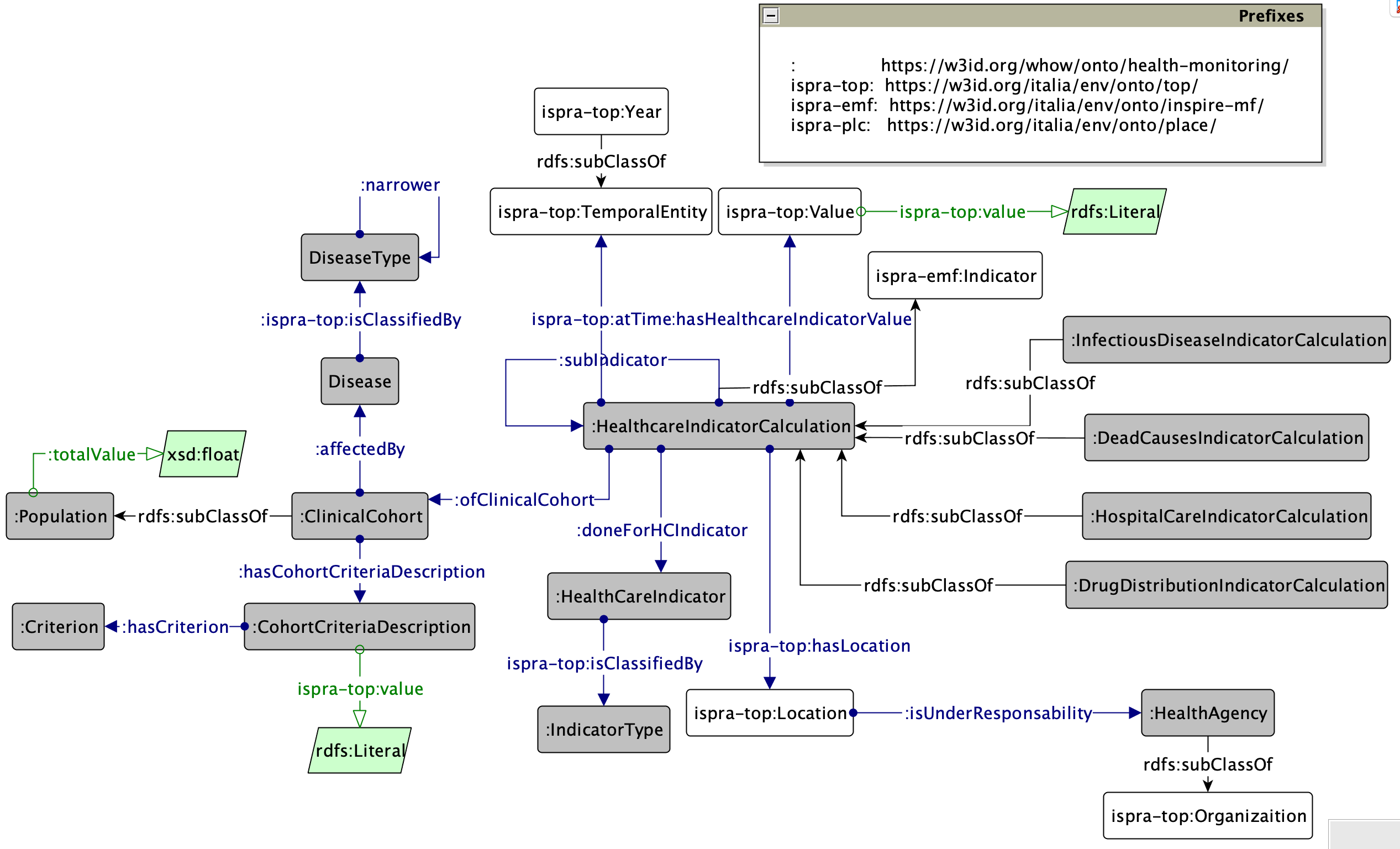} 
\caption{The Health Monitoring ontology.}
\label{fig:hm}
\end{figure}

\paragraph{\textbf{Health Monitoring module.}}
Finally, the {\em Health Monitoring} ontology, whose prefix is \texttt{hm:}\footnote{The prefix \textcode{wh-mon:}, stands for the namespace \url{https://w3id.org/whow/onto/weather-monitoring}.} reuses the OntoPiA Indicator ontology and focuses on the representation of health indicators coming from regional healthcare facilities. Examples include drug distribution rates and hospital accesses according to disease code and facility involved (cf. Figure~\ref{fig:hm}). Different types of \textcode{hm:HealthcareIndicatorCalculation} are defined, based on the typology of indicator they describe, i.e. infectious disease rate, death rates related to diagnosis, average hospital stay and drug distribution. The indicator calculation also refers to a statistical dimension class, \textcode{hm:ClinicalCohort}, which specifies the population referred to as defined by a number of criteria, that is \textcode{hm:CohortCriteriaDescription}, such as age and gender. By reusing the \textcode{ispra-top:} ontology, it is also possible to model the health agency that supervises a specific area.

\subsection{Linked Open Data}
\label{sec:label}
We produced the Linked Open Data from two data providers, i.e. ISPRA and ARIA, as reported in Table~\ref{tab:datasets} by executing the RML mapping as described in Section~\ref{sec:methodology}. Hence, we generated two linked open datasets, that is the one from the data provided by ISPRA and the other from the data provided by ARIA. The ownership of the generated linked datasets along with their corresponding maintenance effort is kept by the data providers. This fits the requirement of WHOW to create and maintain a knowledge graph following a decentralised and distributed paradigm. In this scenario new data providers might publish their data as linked open data compliant with the WHOW ontology network by using their preferred persistent URIs and setting up their own SPARQL endpoint, thus maximising the sustainability of the WHOW-KG. With this regards, ISPRA identified \textcode{https://w3id.org/italia/env/ld/} as its reference namespace. Accordingly, the pattern \texttt{\small{https://w3id.org/italia/env/ld/{\em\{type\}}/{\em \{id\}}}} was applied for producing RDF resources, where \texttt{\small{{\em\{type\}}}} and \texttt{\small{{\em\{id\}}}} are placeholders for an entity type (e.g. \textcode{water-sample}) and its local identifier (e.g. \textcode{45.60555-13.72195}), respectively. The RDF data produced by ISPRA can be queried through their dedicated SPARQL endpoint\footnote{\url{https://dati.isprambiente.it/sparql}.} and are available as a single dump on Zenodo\footnote{\url{https://doi.org/10.5281/zenodo.7916383}} for download with a CC-BY 4.0 International licence.
Similarly, ARIA identified \textcode{https://w3id.org/italia/lombardia/data/} as its reference namespace. Also in this case, the pattern \texttt{\small{https://w3id.org/italia/lombardia/data/{\em\{type\}}/{\em \{id\}}}} was applied for producing RDF resources with the same rationale as before. Again, the RDF data produced by ARIA can be queried via SPARQL\footnote{\url{http://18.102.46.55:18890/sparql}.} and are available on Zenodo\footnote{\url{https://doi.org/10.5281/zenodo.7916732}} for download with a CC0 licence.
Finally, three controlled vocabularies were produced from the data provided by ARIA. This vocabularies provides term definitions for: (i) chemical substances\footnote{\url{https://github.com/whow-project/semantic-assets/blob/main/controlled-vocabularies/chemical-substances/chemical-substances.ttl}.}; (ii) diseases\footnote{\url{https://github.com/whow-project/semantic-assets/blob/main/controlled-vocabularies/diseases/diseases.ttl}}; and (iii) water indicators\footnote{\url{https://github.com/whow-project/semantic-assets/blob/main/controlled-vocabularies/water-indicators/water-indicators.ttl}.}. 
In the case of controlled vocabularies we opted for a namespace not depending on the specific data provider, i.e. \textcode{https://w3id.org/whow/controlled-vocabulary/}. This namespace was used for producing RDF resources by applying the pattern \texttt{\small{https://w3id.org/whow/controlled-vocabulary/{\em\{name\}}/{\em \{id\}}}}, where \texttt{\small{{\em\{name\}}}} and \texttt{\small{{\em \{id\}}}} are placeholders for the vocabulary name (e.g. \textcode{chemical-substances}) and term local identifier (e.g. \textcode{cas-102851-06-9}), respectively. The controlled vocabularies are available on Zenodo\footnote{\url{https://doi.org/10.5281/zenodo.7919460}.} and can be queries via SPARQL\footnote{\url{https://semscout.istc.cnr.it/sparql/}.}. The WHOW-KG counts of 52,943,768 triples in the linked dataset generated by ISPRA, 47,628,449 triples in the linked dataset generated by ARIA, and 16,350 triples available in the controlled vocabularies.

\section{Impact, versioning, and licensing}
\label{sec:evalimpact}
\paragraph{\textbf{Impact.}} The UN Sustainable Development Goal (SDG) no. 6 on clean water and sanitation requires to invest in adequate infrastructure, provide sanitation facilities, and encourage hygiene. 
The importance of considering UN Sustainable Development Goals (SDGs) in the context of open data emerges from several contexts. Notable is the European Parliament resolution of 14 March 2019 on the Annual strategic report on the implementation
and delivery of the SDGs (2018/2279(INI))\footnote{\url{https://eur-lex.europa.eu/legal-content/EN/TXT/HTML/?uri=CELEX:52019IP0220&from=EN}} where a precise call on the Commission is mentioned in order to add data related to the SDGs to the high-value datasets as defined in the directive on open data and public sector information, encouraging the Member States to publish all
reports on the SDGs under a free license. The World Bank Group, in a blog post\footnote{\url{https://blogs.worldbank.org/digital-development/sustainable-development-goals-and-open-data}.} from as far back as 2015, explicitly highlights that ``Open Data can help
achieve the SDGs by providing critical information on natural resources, government operations, public services, and population demographics''. To this end, the WHOW-KG embodies fine-grained thematic indicators that have been identified by data providers and co-creators of the WHOW project according to the three use cases and their legislation bases. We recorded evidences by means of the co-creation programme that the WHOW-KG is of utmost importance to the community encompassing decision makers, practitioners, and data providers in the area of water quality and sanitation. As a matter of fact, 77 individuals contributed to the co-creation programme from different EU countries.

\paragraph{\textbf{Versioning and Licensing.}} The WHOW-KG is under version control on GitHub\footnote{\url{https://github.com/whow-project/semantic-assets}}. The ontology network, controlled vocabularies and linked dataset produced by ISPRA are realeased with a CC-BY 4.0\footnote{\url{https://creativecommons.org/licenses/by/4.0/}.} licence. Instead, the linked dataset produced by ARIA is realeased with a CC0\footnote{\url{https://creativecommons.org/publicdomain/zero/1.0/}} licence.

\section{Related work}
\label{sec:related}

In the context of the monitoring a pillar is the Semantic Sensor Network Ontology (SSN Ontology)\cite{compton_12}. It allows one to represent sensors and observational processes and implements, for the majority of its semantic elements, the ISO 19156 Observations and Measurements (O\&M) standard, used also as reference model in the INSPIRE context. 

Other European projects target water monitoring data models. This is the case of the ODALA\footnote{\url{https://odalaproject.eu/}.} project that created the ODALA Air \& Water application profile\footnote{https://purl.eu/doc/applicationprofile/AirAndWater/Water.}. The profile builds on a core module derived from both O\&M and the SSN Ontology. ODALA presents concepts similar to those defined in the WHOW water monitoring ontology; this creates the prerequisites for a semantic alignment between these knowledge graphs. In the same direction, \cite{wang_20} describes a knowledge-based approach aiming at water quality monitoring and pollution alerting through the proposed Observational Process Ontology (OPO). Similarly, \cite{diaz_22} presents a three-module water quality ontology that combines numerous standards from different domains to obtain a comprehensive approach to the issue. These standards are, among others, GeoSPARQL\footnote{https://www.ogc.org/standard/geosparql/}, the O\&M and SSN cited above, the RDF Data Cube\footnote{https://www.w3.org/TR/vocab-data-cube/} as well as non-ontological resources associated with standards (WaterML\footnote{https://www.ogc.org/standard/waterml/}). At the European level, the European Environmental Agency publishes a Linked Open Data section\footnote{\url{https://www.eea.europa.eu/data-and-maps/daviz/eionet/data/eea-data}.} that comprises data on water quality monitoring. This data is currently under investigation in order to enable possible links with the proposed WHOW knowledge graph.

As far as the health domain is concerned, although it is difficult to find (linked) open data available for the re-use, interesting resources were taken into account when creating the WHOW-KG. In particular, we mention here the Snomed standard\footnote{\url{https://www.snomed.org/}.} for health terms, that has been re-used in order to create proper links with our produced controlled vocabulary on infectious diseases.

In essence, although a variety of works in both domains can be identified, it is still difficult, to the best of our knowledge, to get access to a resource capable of linking the two domains together as we propose with the WHOW-KG.

\section{Conclusions and future work}
\label{sec:conlusions}
In this paper, we have introduced the Water Health Knowledge Graph (WHOW-KG) that links water quality observations with health parameters (e.g. infectious disease rates), thus implementing the well-known connection of water quality effects on people's health. The WHOW-KG is (i) distributed among different data providers, (ii) open to maximise re-use, (iii) multilingual in that labels and comments are provided in both Italian and English, when possible, and (iv) built according to FAIR principles, applied to both ontologies and linked open data.  The WHOW-KG is continuously evolving with further datasets. The aim, in fact, is to provide a resource that can self-sustain and feed itself beyond the duration of the European WHOW project in which it was conceived. In this context, we are planning a number of activities to further increase the visibility of the knowledge graph and its use for any purpose of interest.  Firstly, we are defining SHACL shapes, starting from the OWL restrictions defined in the ontology network, to support the overall validation phase of the proposed methodology. Secondly, in order to open ourselves up to a wider audience of possible developers, part of our future work is to define rest APIs based on the semantics defined through the ontology network. Thirdly, in order to maximise the possibilities of re-use in a wider European context, we will exploit services such as eTranslation\footnote{\url{https://commission.europa.eu/resources-partners/etranslation}.} to provide additional languages for datasets and ontologies, making the knowledge graph understandable to possible stakeholders from different European countries. Finally, the knowledge graph will be made available through a series of national and European platforms. In fact, we plan to publish the linked open datasets in the Italian national catalogue of open data, thanks to the implementation of the DCAT-AP metadata profile, and from there to \texttt{data.europa.eu}. As for the ontologies we are planning to require their inclusion in the Italian national catalogue of semantic assets named schema.gov.it.

\section*{Acknowledgements}
This work has been supported by the Water Health Open knoWledge (WHOW) project co-financed by the Connecting European Facility programme of the European Union under grant agreement INEA/CEF/ICT/A2019/206322.

\printbibliography

@article{compton_12,
  author = {Compton, Michael and Barnaghi, Payam M. and Bermudez, Luis and Garcia-Castro, Raul and Corcho, {\'O}scar and Cox, Simon J. D. and Graybeal, John and Hauswirth, Manfred and Henson, Cory A. and Herzog, Arthur and Huang, Vincent A. and Janowicz, Krzysztof and Kelsey, W. David and Phuoc, Danh Le and Lefort, Laurent and Leggieri, Myriam and Neuhaus, Holger and Nikolov, Andriy and Page, Kevin R. and Passant, Alexandre and Sheth, Amit P. and Taylor, Kerry},
  journal = {Journal of Web Semantics},
  pages = {25--32},
  title = {The {SSN} ontology of the {W3C} semantic sensor network incubator group},
  volume = 17,
  year = 2012,
  doi = {10.1016/j.websem.2012.05.003}
}

@article{diaz_22,
  title={Characterizing water quality datasets through multi-dimensional knowledge graphs: a case study of the Bogota river basin},
  author={Juan D. Rond{\'o}n D{\'i}az and Luis Manuel VILCHES-BL{\'A}ZQUEZ},
  journal={Journal of Hydroinformatics},
  year={2022},
  doi = {10.2166/hydro.2022.070}
}

@article{wang_20,
author = {Wang, Xiaolei and Wei, Haitao and Chen, Nengcheng and He, Xiaohui and Tian, Zhihui},
year = {2020},
month = {03},
pages = {715},
title = {An Observational Process Ontology-Based Modeling Approach for Water Quality Monitoring},
volume = {12},
journal = {Water},
doi = {10.3390/w12030715}
}

@inproceedings{SSOPattern,
  author = {Janowicz, Krzysztof and Compton, Michael},
  booktitle = {SSN 2010},
  editor = {Taylor, Kerry and Ayyagari, Arun and Roure, David De},
  ee = {http://ceur-ws.org/Vol-668/paper12.pdf},
  publisher = {CEUR-WS.org},
  series = {CEUR Workshop Proceedings},
  title = {{The Stimulus-Sensor-Observation Ontology Design Pattern and its Integration into the Semantic Sensor Network Ontology}},
  volume = 668,
  year = 2010
}

@book{inspire,
author = {Cox, Simon},
year = {2011},
month = {01},
pages = {1--32},
title = {Guidelines for the use of Observations \& Measurements and Sensor Web Enablement-related standards in INSPIRE Annex II and III data specification development},
publisher = {
INSPIRE Maintenance and Implementation Group (MIG)}
}

@report{Picone2021,
  author       = {Marco Picone and
                  Gianluca Carletti and
                  Carmen Ciciriello and
                  Paola Esposito},
  title        = {Use Cases Definition},
  month        = dec,
  year         = 2021,
  note         = {{Deliverable n. 2.1 Activity title: Requirements 
                   definition}},
  publisher    = {Zenodo},
  doi          = {10.5281/zenodo.6685761},
  url          = {https://doi.org/10.5281/zenodo.6685761}
}

@report{Carletti2023,
  author       = {Carletti, Gianluca and
                  Picone, Marco and
                  De Angelis, Roberta and
                  Spada, Emanuela and
                  Borrello, Patrizia},
  title        = {{Legal and quality assessment of datasets 
                   identified}},
  month        = mar,
  year         = 2023,
  publisher    = {Zenodo},
  doi          = {10.5281/zenodo.7900842},
  url          = {https://doi.org/10.5281/zenodo.7900842}
}

@article{Carriero2021,
	title = {Pattern-based design applied to cultural heritage knowledge graphs},
	volume = {12},
	issn = {2210-4968},
	doi = {10.3233/SW-200422},
	number = {2},
	journal = {Semantic Web},
	author = {Carriero, Valentina Anita and Gangemi, Aldo and Mancinelli, Maria Letizia and Nuzzolese, Andrea Giovanni and Presutti, Valentina and Veninata, Chiara},
	year = {2021},
	note = {Publisher: IOS Press},
	pages = {313--357},
}

@incollection{Blomqvist2016,
  author = {Blomqvist, Eva and Hammar, Karl and Presutti, Valentina},
  booktitle = {Ontology Engineering with Ontology Design Patterns},
  editor = {Hitzler, Pascal and Gangemi, Aldo and Janowicz, Krzysztof and Krisnadhi, Adila and Presutti, Valentina},
  isbn = {978-1-61499-676-7},
  pages = {23--50},
  publisher = {IOS Press},
  series = {Studies on the Semantic Web},
  title = {{Engineering Ontologies with Patterns - The eXtreme Design Methodology}},
  volume = 25,
  year = 2016,
  doi = {10.3233/978-1-61499-676-7-23}
}

@incollection{Gangemi2009,
  author = {Gangemi, Aldo and Presutti, Valentina},
  booktitle = {Handbook on Ontologies},
  pages = {221--243},
  publisher = {Springer},
  title = {Ontology design patterns},
  year = 2009,
  doi = {10.1007/978-3-540-92673-3_10}
}

@incollection{Gruninger1995,
  title={The role of competency questions in enterprise engineering},
  author={Gr{\"u}ninger, Michael and Fox, Mark S},
  booktitle={Benchmarking—Theory and practice},
  pages={22--31},
  year={1995},
  publisher={Springer}
}

@inproceedings{Blomqvist2010,
  title={Experimenting with eXtreme design},
  author={Blomqvist, E. and Presutti, V. and Daga, E. and Gangemi, A.},
  booktitle={Proceedings of the 17th International Conference on Knowledge Engineering and Knowledge Management},
  pages={120--134},
  year={2010},
  doi={10.1007/978-3-642-16438-5_9},
  editor={Cimiano, P. and Pinto, H.S.},
  series={Lecture Notes in Computer Science},
  volume={6317},
  organization={Springer}
}

@inproceedings{Presutti2016,
  title={The role of ontology design patterns in linked data projects},
  author={Presutti, Valentina and Lodi, Giorgia and Nuzzolese, Andrea and Gangemi, Aldo and Peroni, Silvio and Asprino, Luigi},
  booktitle={Proceedings of the 35th International Conference on Conceptual Modeling},
  editor = {Comyn-Wattiau, Isabelle and Tanaka, Katsumi and Song, Il-Yeol and Yamamoto, Shuichiro and Saeki, Motoshi},
  pages={113--121},
  year={2016},
  doi={10.1007/978-3-319-46397-1_9},
  series = {Lecture Notes in Computer Science},
  volume = {9974},
  organization={Springer}
}

@incollection{Carriero2020,
  author       = {Carriero, Valentina Anita and
                  Daquino, Marilena and
                  Gangemi, Aldo and
                  Nuzzolese, Andrea Giovanni and
                  Peroni, Silvio and
                  Presutti, Valentina and
                  Tomasi, Francesca},
  editor       = {Cota, Giuseppe and
                  Daquino, Marilena and
                  Pozzato, Gian Luca},
  title        = {{The Landscape of Ontology Reuse Approaches}},
  booktitle    = {Applications and Practices in Ontology Design, Extraction, and Reasoning},
  series       = {Studies on the Semantic Web},
  volume       = {49},
  pages        = {21--38},
  publisher    = {{IOS} Press},
  year         = {2020},
  doi          = {10.3233/SSW200033}
}

@article{Haller2019,
  author = {Haller, Armin and Janowicz, Krzysztof and Cox, Simon J. D. and Lefrancois, Maxime and Taylor, Kerry and Phuoc, Danh Le and Lieberman, Joshua and Garcia-Castro, Ra{\'u}l and Atkinson, Rob and Stadler, Claus},
  journal = {Semantic Web},
  number = 1,
  pages = {9--32},
  title = {The modular SSN ontology: A joint W3C and OGC standard specifying the semantics of sensors, observations, sampling, and actuation.},
  volume = 10,
  year = 2019,
  doi = {10.3233/SW-180320}
}

@inproceedings{Carriero2019,
  author = {Carriero, Valentina Anita and Mariani, Fabio and Nuzzolese, Andrea Giovanni and Pasqual, Valentina and Presutti, Valentina},
  booktitle = {ISWC Satellites},
  editor = {Su{\'a}rez-Figueroa, Mari Carmen and Cheng, Gong and Gentile, Anna Lisa and Gu{\'e}ret, Christophe and Keet, C. Maria and Bernstein, Abraham},
  pages = {221--224},
  publisher = {CEUR-WS.org},
  series = {CEUR Workshop Proceedings},
  title = {{Agile Knowledge Graph Testing with TESTaLOD}},
  volume = 2456,
  year = 2019
}

@inproceedings{Dimou2014,
  author = {Dimou, Anastasia and Sande, Miel Vander and Colpaert, Pieter and Verborgh, Ruben and Mannens, Erik and de Walle, Rik Van},
  booktitle = {LDOW},
  editor = {Bizer, Christian and Heath, Tom and Auer, S{\"o}ren and Berners-Lee, Tim},
  publisher = {CEUR-WS.org},
  series = {CEUR Workshop Proceedings},
  title = {RML: A Generic Language for Integrated RDF Mappings of Heterogeneous Data.},
  volume = 1184,
  year = 2014
}

@inproceedings{Falco2014,
  author = {Falco, Riccardo and Gangemi, Aldo and Peroni, Silvio and Shotton, David M. and Vitali, Fabio},
  booktitle = {Satellite Events of ESWC},
  editor = {Presutti, Valentina and Blomqvist, Eva and Troncy, Rapha{\"e}l and Sack, Harald and Papadakis, Ioannis and Tordai, Anna},
  isbn = {978-3-319-11954-0},
  pages = {320--325},
  publisher = {Springer},
  series = {Lecture Notes in Computer Science},
  title = {{Modelling OWL Ontologies with Graffoo}},
  volume = 8798,
  year = 2014,
  doi = {10.1007/978-3-319-11955-7_42}
}

\end{document}